\documentclass[amsmath,prb,showpacs]{revtex4}
\usepackage{graphicx}
\usepackage{booktabs}

\begin{document}

\title{Properties of low density quantum fluids within nanopores}
\author{Nathan M. Urban}
\email{nurban@psu.edu}
\affiliation{Department of Physics, Pennsylvania State University, University Park, Pennsylvania 16802, USA}
\altaffiliation[Current address: ]{Department of Geosciences, Pennsylvania State University, University Park, Pennsylvania 16802, USA}
\author{Milton W. Cole}
\email{miltoncole@aol.com}
\affiliation{Department of Physics, Pennsylvania State University, University Park, Pennsylvania 16802, USA}
\author{E. Susana Hern\'andez}
\email{shernand@df.uba.ar}
\affiliation{Departamento de F{\'\i}sica, Facultad de Ciencias Exactas y
Naturales, Universidad de Buenos Aires, Argentina}
\affiliation{Consejo Nacional de Investigaciones Cientficas y T\'ecnicas, Argentina}

\begin{abstract}
The behavior of quantum fluids ($^4$He and H$_2$) within nanopores
is explored in various regimes, using several different methods. A
focus is the evolution of each fluid's behavior as pore radius $R$
is increased. Results are derived with the path integral Monte Carlo
method for the finite temperature ($T$) behavior of quasi-one-dimensional
(1D) liquid $^4$He and liquid H$_2$, within pores of varying $R$.
Results are also obtained, using a density functional method, for
the $T=0$ behavior of $^4$He in pores of variable $R$.
\end{abstract}


\maketitle

\section{Introduction}
\label{sec:intro}

The behavior of quantum fluids (e.g. helium or hydrogen) in reduced
dimensionality is one of the most interesting topics in low temperature
physics.\cite{ref1} Among the systems of particular interest are
monolayer films (which are effectively two-dimensional, or 2D) and
fluids within narrow pores (1D) or droplets
(0D).\cite{ref2,ref3,ref4,ref5,ref6,ref7} Properties associated
with phase transitions (e.g. critical exponents) are particularly
sensitive to the dimensionality, as exemplified by the Kosterlitz-Thouless
theory of the 2D superfluid transition in He films.\cite{ref8}
Confined ideal gases are particularly amenable to theoretical study,
because one need solve just the single particle Schr\"odinger
equation in the given environment, rather than the fully interacting
many-body problem.\cite{ref7}  Such an idealized model can provide
qualitatively, and sometimes quantitatively, reliable predictions
of the effects of confinement on the fluid. Furthermore, these
calculations can be supplemented, in some cases, by virial expansions
in order to assess the role of interactions, presuming their effects
to be weak.\cite{ref9,ref10}  In contrast, dense quantum fluids
require the use of the path integral Monte Carlo (PIMC) or diffusion
Monte Carlo methods, which are computationally demanding approaches.
Nevertheless, these techniques has been applied in recent years to
many problems of reduced dimensionality.\cite{ref11,ref12}

Considerable progress in understanding $D<3$ behavior has also been
achieved on the experimental front. Although the evolution of
superfluidity in thin films was first observed in heat capacity
experiments nearly 70 years ago, that subject remains incompletely
understood.\cite{ref13}  The problem of quantum droplets has received
particular attention in recent years, stimulated especially by the
search for superfluid hydrogen.\cite{ref6,ref14,ref15}  The discovery
of carbon nanotubes and interest in other quasi-1D porous media
have led to many investigations of 1D quantum
fluids.\cite{ref5,ref16,ref17}

This paper addresses a set of problems involving helium and hydrogen
within small radius ($R$) cylindrical pores, such as carbon nanotubes.
For $R\sim 0$, one encounters the strictly 1D limit of a quantum
fluid.  While there has been considerable exploration of the zero
temperature ($T$) behavior of such systems, relatively few studies
of finite $T$ behavior of the interacting system have been carried
out.\cite{ref12,ref17,ref18,ref19,ref20,ref21,ref34}  In a recent
study, we explored such behavior for 1D $^4$He.  It was demonstrated
that a 1D phonon model, analogous to the Landau hydrodynamic theory
of liquid $^4$He, does describe the behavior at low $T$ of 1D $^4$He
at moderate to high density (compared to the equilibrium density
$\rho=0.036$/\AA).\cite{ref22}  Section \ref{sec:4he-0T} of the
present paper presents a finite range density functional study of
the behavior of $^4$He, at temperature $T=0$, as $R$ is varied.
The results exhibit the evolution of the system from an essentially
1D ``axial phase'' to a ``cylindrical shell'' phase, localized near
the pore walls.\cite{ref16,ref23}  Section \ref{sec:h2-finiteT}
shows how the behavior of H$_2$ at finite $T$, but low density,
varies with pore radius $R$.  Section \ref{sec:summary} summarizes
our results.

\section{Liquid $^4$He in a nanopore at zero temperature}
\label{sec:4he-0T}

As shown in previous work, finite-range density functional (FRDF)
calculations permit the calculation of the equation of state (EOS)
of helium atoms in various environments (see, e.g., Barranco
\textit{et~al.}\cite{ref15} and references therein). In this section,
we investigate the condensation of helium in a cylindrical pore of
radius $R$ and length $L$, with the same density functional employed
in Refs.~\onlinecite{ref24,ref25,ref26}, with detailed description
and parameters found in Ref.~\onlinecite{ref27}. The calculation
is based on the energy density $\mathcal{E}[\rho(r)]$, where $\rho(r)$
is the particle density, normalized to the total number of helium
atoms $N$ so that, in the pore geometry, the linear density (over
$2\pi$), $n = N/(2\pi L)$, is
\begin{equation}
n =  \int dr\,r \rho(r)\,.\label{eq:1}
\end{equation}

The total energy $e = E/(2\pi L)$ per unit length (over $2\pi$) is then
\begin{equation}
e =  \int dr\,r \mathcal{E}[\rho(r)]\,.\label{eq:2}
\end{equation}

The Euler-Lagrange equation for the density is obtained by functional
differentiation of this energy with respect to $\rho(r)$, at fixed
total number of particles, and takes the form
\begin{equation}
\Bigl[-\frac{\hbar^2}{2m}\Bigl(\frac{d^2}{dr^2}+\frac{1}{r}\frac{d}{dr}\Bigr)+U(\rho)+V(r;R)\Bigr]\sqrt{\rho(r)} = \mu\sqrt{\rho(r)}\,,\label{eq:density}
\end{equation}
with $\mu$ the chemical potential, $U(\rho)\equiv \delta
E/\delta\rho(r)$ the density-dependent mean field and 
$V(r;R)$ the adsorption potential provided by the cylindrical sheet.
This function was derived by considering that a He atom at radial
distance $r$ from the axis of the cylinder interacts pairwise with
continuum sheet of atoms, of uniform areal density $\theta$, via a
Lennard-Jones (LJ) potential of well-depth $\epsilon$ and hard core
diameter $\sigma$.  The result is expressed in terms of elliptic
integrals defined in Ref.~\onlinecite{ref23}:
\begin{equation}
V(r;R) = 3\pi\epsilon\sigma^2\theta\Bigl[\frac{21}{32}\Bigl(\frac{\sigma}{R}\Bigr)^{10} M_{11}(r/R) - \Bigl(\frac{\sigma}{R}\Bigr)^4 M_5(r/R)\Bigr]\,.\label{eq:3}
\end{equation}
Here $M_{2n+1}$ is an elliptic integral over angle $\phi$.

Prior to determining the EOS of liquid He in pores, we consider the
binding of a single He atom to an alkali metal cylindrical sheet.
In particular, Cs substrates of diverse shapes are excellent
candidates to investigate nontrivial aspects of the physics of
wetting. While the He-Cs potential is too weakly attractive to
adsorb He on a flat Cs surface (below saturation chemical potential
$\mu_0$), there are two reasons why He does adsorb within a pore.
One is that the surface tension cost of pore-filling is smaller
than that on a planar surface (by a factor of two\cite{ref28}). The
other is that the potential can be substantially enhanced within a
concave environment, either a pore or a wedge of intermediate
openings.\cite{ref27}

\begin{figure}
\includegraphics[width=12cm]{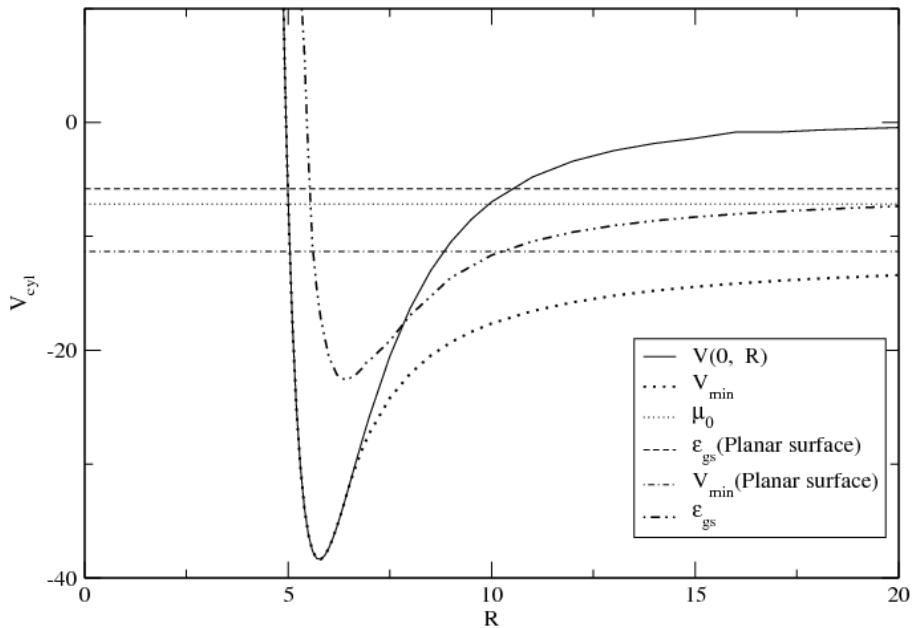}
\caption{\label{fig1}
Cylindrical potential $V(0,R)$ on the axis, potential $V_\mathrm{min}$
at the absolute minimum, and ground state energy $\varepsilon_{gs}$
of a $^4$He atom. For reference, the bulk chemical potential $\mu_0$,
the minimum potential of a planar Cs sheet $V_\mathrm{min}$ (planar
surface) and corresponding energy $\varepsilon_{gs}$ (planar surface)
are also displayed.}
\end{figure}

\begin{figure}
\includegraphics[width=10cm]{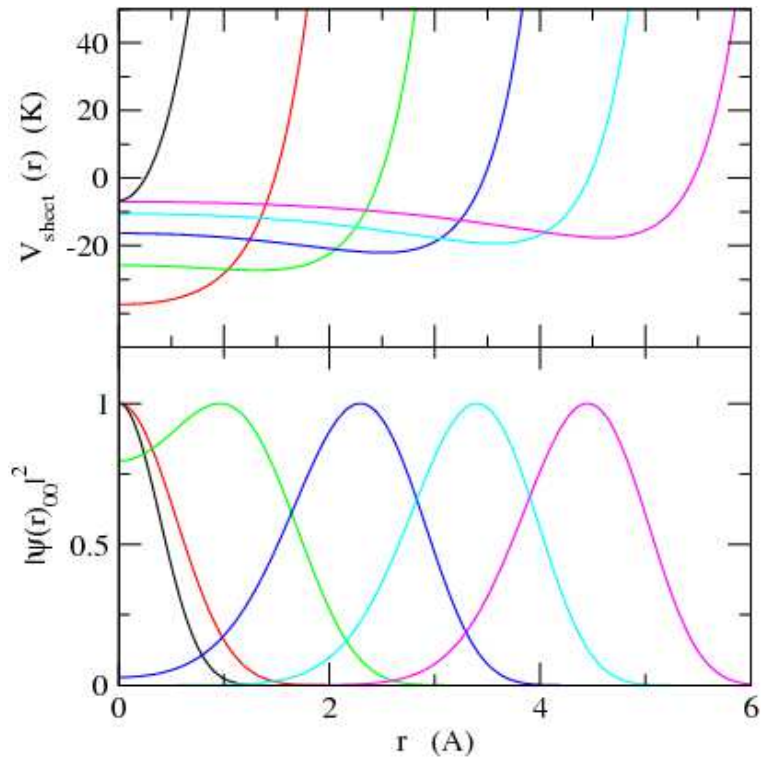}
\caption{\label{fig2}
\textit{Upper panel:} potentials of a He atom inside Cs cylinders
of radius for $R$ values between 5 and 10~\AA, from left to right
in steps of 1~\AA. \textit{Lower panel:} corresponding probability
density of one atom in its ground state. The 3D probability
distributions have been normalized to unit maximum value.}
\end{figure}

Illustrative results are shown in Figs.~\ref{fig1} and \ref{fig2}.
In Fig.~\ref{fig1} we plot, as functions of cylinder radius $R$,
the potential $V(0,R)$ on the cylinder axis, the potential
$V_\mathrm{min}$ at the absolute minimum of $V(r,R)$, and the ground
state (gs) energy $\varepsilon_{gs}$ of one atom obtained by solving
the single-particle (sp) Schr\"odinger equation in the cylindrically
symmetric potential.  The $^4$He bulk chemical potential $\mu_0 =
-7.15$~K is displayed as a reference, as well as the minimum potential
$V_{\mathrm{min}\infty} = -11.3$~K of a planar Cs sheet and the
corresponding gs energy $\varepsilon_{gs\infty} = -5.83$~K.  We note
that the absolute minimum departs from the axis at a radius of about
6.8~\AA. On the other hand, the sp energy drops below the potential
on the axis around $R = 7.5$~\AA, corresponding to an approximate
threshold radius for the off-axis displacement of the maximum
probability density $\lvert\psi(r)\rvert^2$.  This behavior is illustrated
in Fig.~\ref{fig2}, where the upper panel displays the potentials
due to Cs sheets of varying radius $R$, as functions of the radial
distance $r$, and the lower panel shows the probability density of
the He atom.  The parameters of the He-Cs interaction are
$\epsilon=2.795$~K, $\sigma=5.31$~\AA, and $\theta=0.38$/\AA$^2$.
The large value of $\sigma$ reflects the extent of the hard-core
repulsion for this interaction, which is responsible for the weak
van der Waals attraction, i.e., the small value of $\epsilon$.

For the smallest radii displayed, 5 and 6~\AA, the probability
density for this ``axial phase'' is seen to be concentrated on the
axis, while for the largest radii, 9 and 10~\AA, the particle is
localized (with spread $\sim 1.2$~\AA) about 5.6~{\AA} from the
wall; for this ``cylindrical shell'' phase, the latter distance
asymptotes to $\sigma=5.31$~{\AA} for the hypothetical limiting
case $R=\infty$.  In the intermediate $R$ regime, there remains a
non-negligible probability density at $r=0$, seen to be quite sizable
for $R = 7$~\AA. Note that the atom is unbound (relative to vacuum)
in a 5~{\AA} tube; in fact, according to Fig.~\ref{fig1}, the gs
energy is negative for radii larger than 5.2~\AA. Thus, a noninteracting
Bose gas will be bound within such pores in a quasi-1D axial phase,
a filled pore regime and a quasi-2D shell phase, as $R$ increases.
For even larger $R$, the usual layering structure of helium in a
cavity occurs.\cite{ref16,ref26}

\begin{figure}
\includegraphics[width=12cm]{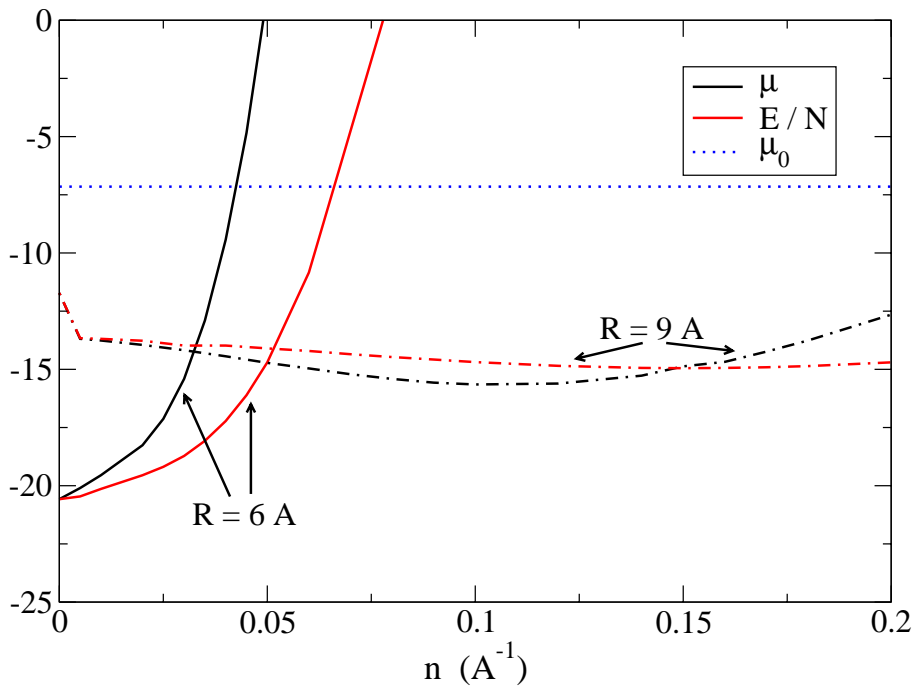}
\caption{\label{fig3}
Chemical potential (black) and energy per particle (red) for liquid
He in Cs pores of radii $R = 6$ and 9~\AA, as functions of linear
density.  The bulk chemical potential $\mu_0$ is shown as a dotted
line.}
\end{figure}

\begin{figure}
\includegraphics[width=12cm]{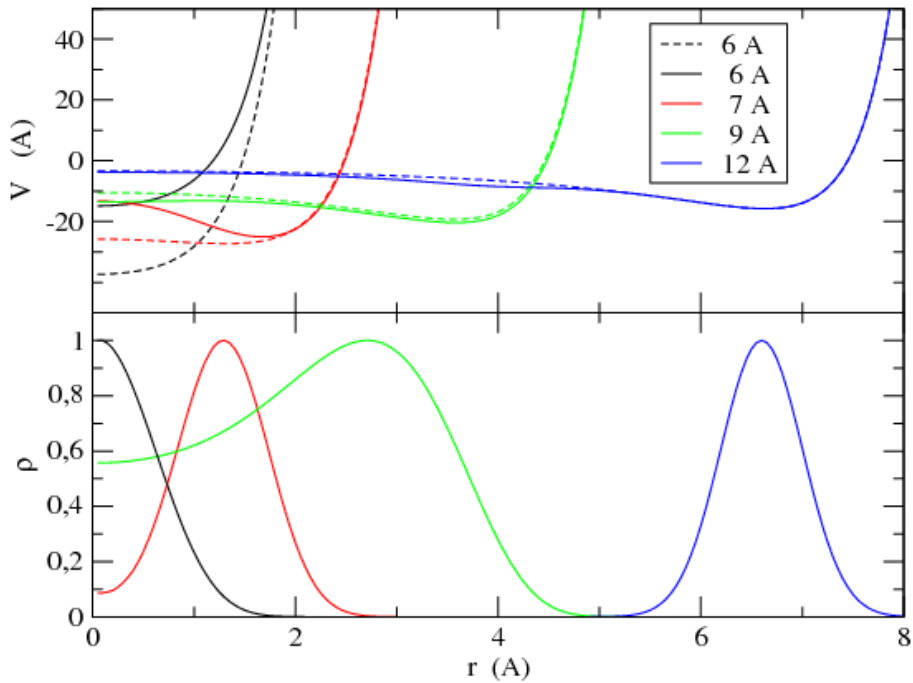}
\caption{\label{fig4}
\textit{Upper panel:} total mean-field potential in Eq.~\ref{eq:density}
(full lines) and potentials $V(r,R)$ from the Cs cylinder for radii
$R = 6$, 7, 9 and 12~{\AA} (from left to right). \textit{Lower
panel:} corresponding particle densities (normalized to unit maximum)
at a linear density $n = 0.05/$\AA.}
\end{figure}

Next, we explore the many-body behavior by solving Eq.~\ref{eq:density}
to obtain the chemical potential $\mu$, the total energy, and the
density profiles $\rho(r)$, as functions of the linear density
$n$.  Significant results are presented in Fig.~\ref{fig3}, which
shows $\mu$ and the energy per particle for two selected pore radii,
namely $R=6$ and 9~\AA, in full and dashed lines, respectively.
For the narrower pore, a stable quasi-one-dimensional (Q1D) system
with negative grand potential per unit length, $\omega = e-\mu$,
is found for any linear density below $0.042$~\AA$^{-1}$, above
which the system becomes unstable. For a sufficiently wide pore,
the particle distribution has moved towards the cylinder wall and
the EOS displays the typical 2D prewetting jump at $n_{pw}=
0.15$~\AA$^{-1}$, at which point the energy per particle becomes
identical to the chemical potential, resulting in condensation of
a stable film at this or higher linear densities. This behavior can
be assessed by examining density profiles for various values of $R$
at a given linear density, as displayed in Fig.~\ref{fig4} for $n
= 0.05$~\AA$^{-1}$.  In this figure, the upper panel shows both the
confining potential and the total one-body field $U[\rho(r)]+V(r;R)$
experienced by particles in the fluid.  The resulting particle
densities are plotted in the lower panel as functions of distance
$r$.

At first sight, in Fig.~\ref{fig4} we note the presence of the three
distinct regimes of density behavior, as encountered in the sp
pattern displayed in Fig.~\ref{fig2}.  In addition, this plot
provides evidence of the effect of cohesion among the He atoms and
its competition with the weak adhesion to the confining walls. The
difference between full and dashed lines is the total interaction
energy (per particle) of the interacting fluid, as shown in various
prior studies.\cite{ref24,ref25,ref26,ref27}  This total consists
of a screened LJ interaction plus two density-dependent correlation
terms, one attractive, the other repulsive, representing the effects
of the many-body environment on the one-body field. For the two
smallest radii, the overall outcome, as seen in the upper panel of
Fig.~\ref{fig4}, is a visible reduction in potential energy and a
distortion of the potential well that causes an earlier departure
of the minimum off-axis, as compared with the sp picture in
Fig.~\ref{fig2}.  The situation is reversed for the largest radii.
For $R = 9$~\AA, the potential energy gain of about 2~K near the
cylinder axis is sufficient to sustain a non-negligible amount of
fluid in that region. As a consequence, pores of radii between
somewhat less than 7~{\AA} and more than 9~{\AA} result in a particle
density spread out from axis to wall, and the appearance of the
purely axial regime shifts towards larger radii. For radii near
12~{\AA} and above, instead, the correlation effects in the helium
fluid are negligible and the system behaves essentially as a confined
gas.

\section{H$_2$ in a nanopore at finite $T$}
\label{sec:h2-finiteT}

We consider in this section the adsorption of H$_2$ (which, after
$^4$He, is the textbook quantum fluid) molecules inside a cylindrical
pore within a bulk solid. As in the preceding section, a low density
gas of molecules within a large pore will typically adsorb first
onto the inner surface of the pore, creating a 2D cylindrical shell
phase at low densities; at higher densities, central regions of the
pore will fill with a 3D bulk adsorbate.\cite{ref16} Very narrow
pores, however, will constrain the adsorbate to lie near the pore
axis, forming a quasi-1D (``axial'') phase. This variation of
behavior provides the opportunity to control the effective
dimensionality of the adsorbate by adjusting the size of the pore
into which it adsorbs. For concreteness, we chose to study pores
in MgO glass, a much more strongly adsorbing substrate than the Cs
discussed in the previous section; the well-depth (48~meV) of an
H$_2$ molecule on planar MgO is about 15 times higher than that on
Cs.~\cite{ref29,ref30} MgO nanopores can be fabricated as small as
3~nm in diameter.\cite{ref43}  The H$_2$-MgO potential was taken
to be derived from Lennard-Jones 12-6 pair interactions, integrated
over the bulk MgO solid; these gas-surface interaction parameters
are given by $\sigma=2.014$~\AA, $\epsilon=45.91$~K. The H$_2$-H$_2$
interaction potential was taken to be of the Silvera-Goldman
form.\cite{ref31} The molecule is treated as a spherically symmetric
point particle, neglecting internal degrees of freedom.

\subsection{Structure: path integral Monte Carlo simulations}

This quantum statistical system can be simulated by the path integral
Monte Carlo (PIMC) method.\cite{ref11,ref33} The simulations were
performed for various pore radii $R$ at fixed temperature $T=0.5$~K
(or 1~K) and particle number $N=28$; the mean 3D density $\bar\rho=N/V$
changed as the pore volume was varied. The value $N=28$ was chosen
to approximate the equilibrium linear density of a 1D system of
fixed length (of the periodic unit cell) $L=125$~\AA.\cite{ref34,ref35}
The value of $L$ was chosen large enough to ensure that finite size
effects were minimal, as confirmed by simulation tests.\cite{ref36}

\begin{figure}
\includegraphics[width=10cm,angle=-90]{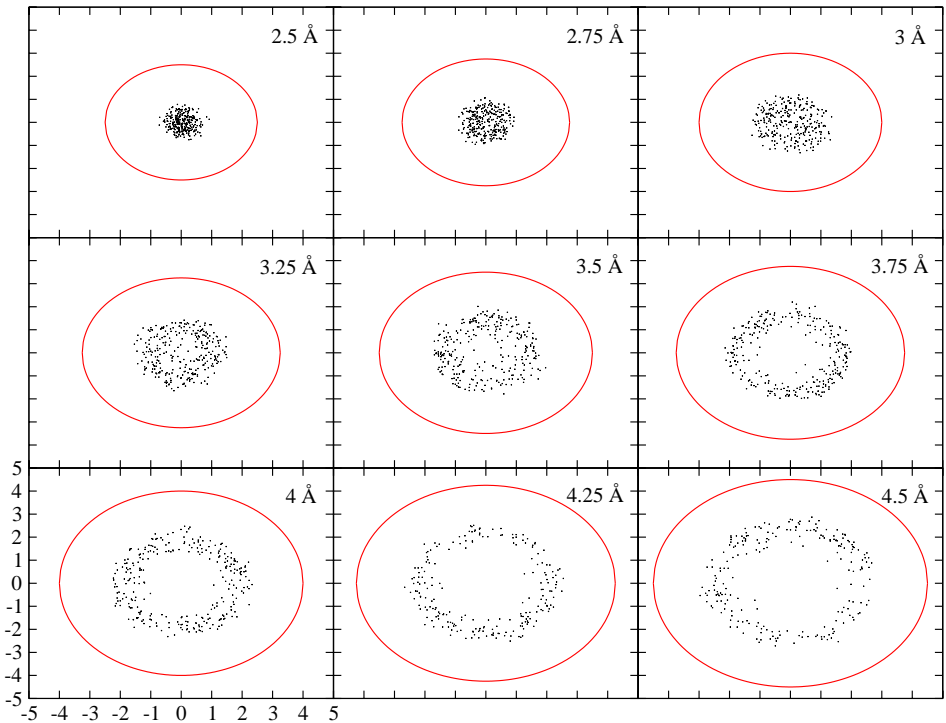}
\caption{\label{fig:h2-pore-density}
H$_2$ density distribution [at $T=0.5$~K, $\bar\rho = (2.55\times
10^{-3}$~\AA)$/R^2$], viewed in transverse section, for pore radii
$R = 2.5$--$4.5$~\AA, as indicated. Depicted are several hundred
superimposed Monte Carlo samples of particle positions (dots) for
a system of $N=28$ particles, along with the inner surface of the
pore (circles). (The circles are distorted into ellipses due
to the aspect ratio of the figure.) The H$_2$ evolves from a 1D
(axial) to a 2D (cylindrical shell) phase as $R$ increases.}
\end{figure}

\begin{figure}
\includegraphics[width=10cm,angle=-90]{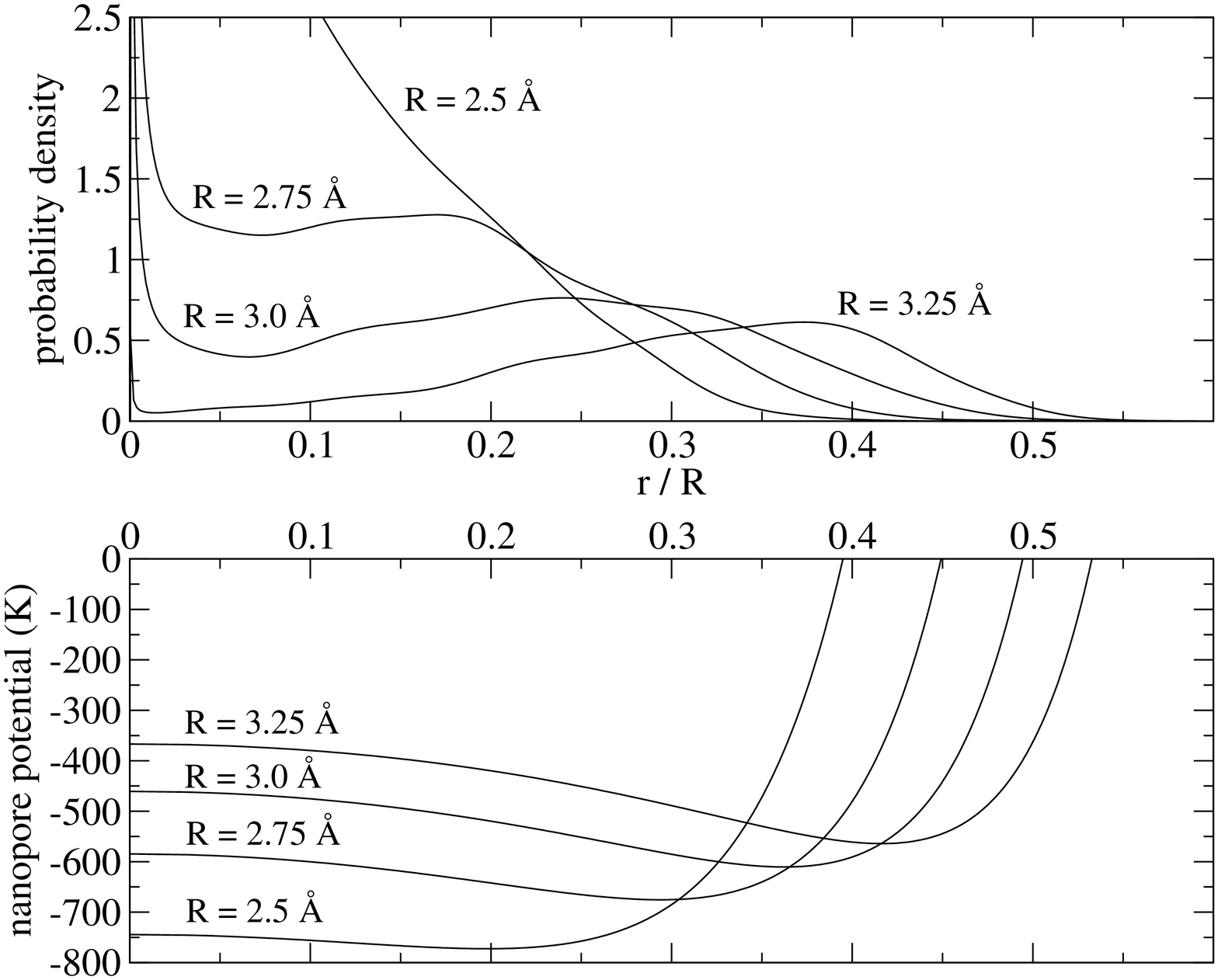}
\caption{\label{fig:radial-pdf}
H$_2$ probability density [at $T=0.5$~K, $\bar\rho=(2.55\times
10^{-3}$~\AA)/$R^2$] and MgO pore potential, for pore radii $R=2.5$,
2.75, 3.0 and 3.25~\AA, as functions of dimensionless radius $r/R$.
(Radial densities near $r=0$ are exaggerated due to finite size
effects after normalizing the radial distribution by $1/(2\pi r)$
to obtain the probability density.)}
\end{figure}

The distribution of H$_2$ molecules calculated by PIMC simulation
is given in Fig.~\ref{fig:h2-pore-density} for several pore sizes.
The axial-to-shell transition is more apparent in
Figure~\ref{fig:radial-pdf}, which shows the H$_2$ radial density
distribution, along with the corresponding H$_2$-pore potential
energy curves. Classically, the molecules will concentrate on axis
when the potential minimum is at $r=0$, and will move off axis when
the minimum is found at some $r>0$, if intermolecular interactions
are neglected (the low density limit).  The PIMC simulations at low
densities indicate that the adsorbate moves off axis between pore
radii of $R=2.5$ and 2.75~\AA; by comparison, the classical potential
minimum moves off axis at slightly smaller pore size, $R=2.35$~\AA.
In other words, the molecules tend to concentrate near the axis
more when treated as quantum particles than when treated classically.
The comparison between quantum and classical behavior in this system
is discussed more fully in the following section.

\begin{figure}
\includegraphics[width=6cm]{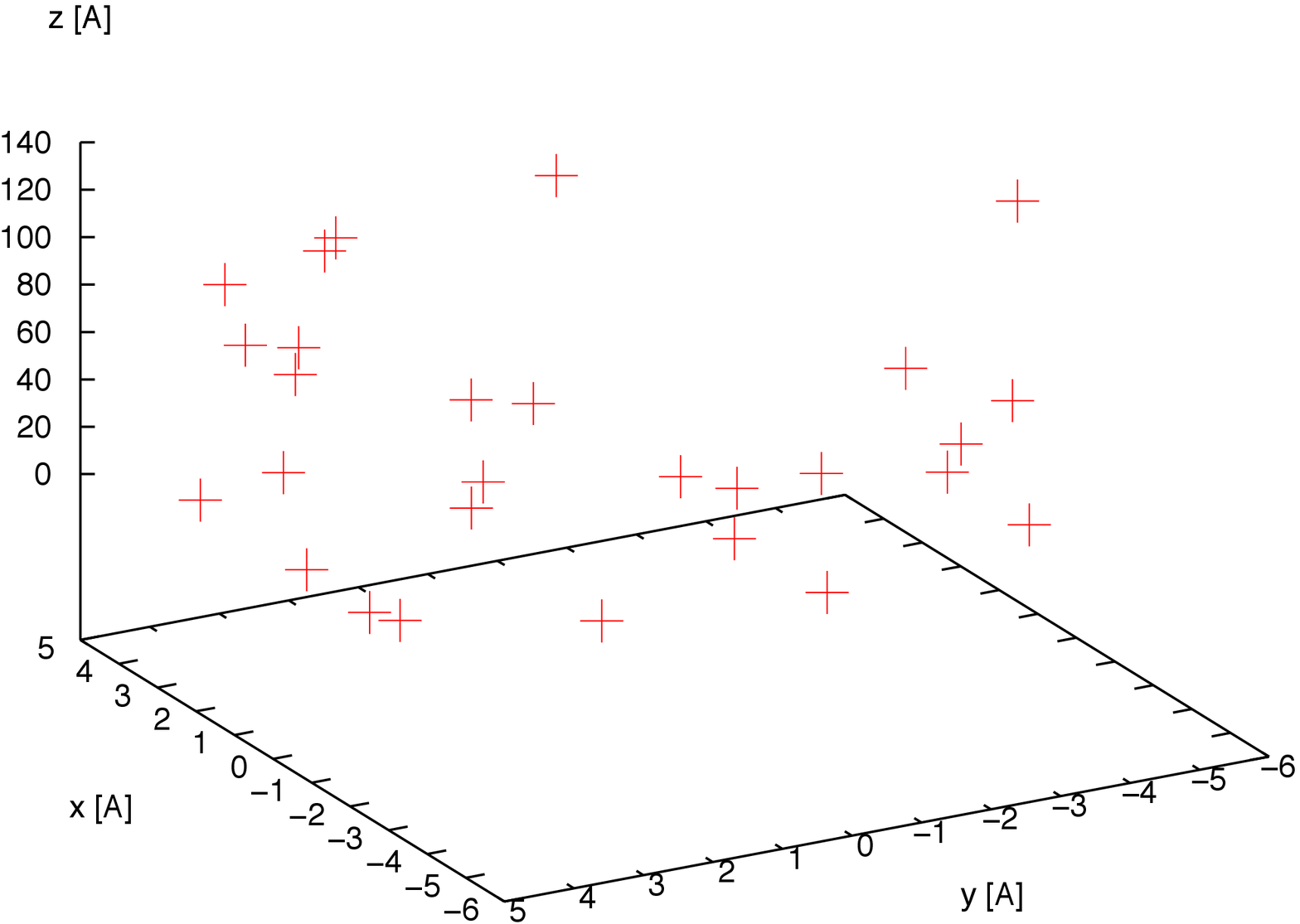}
\includegraphics[width=6cm]{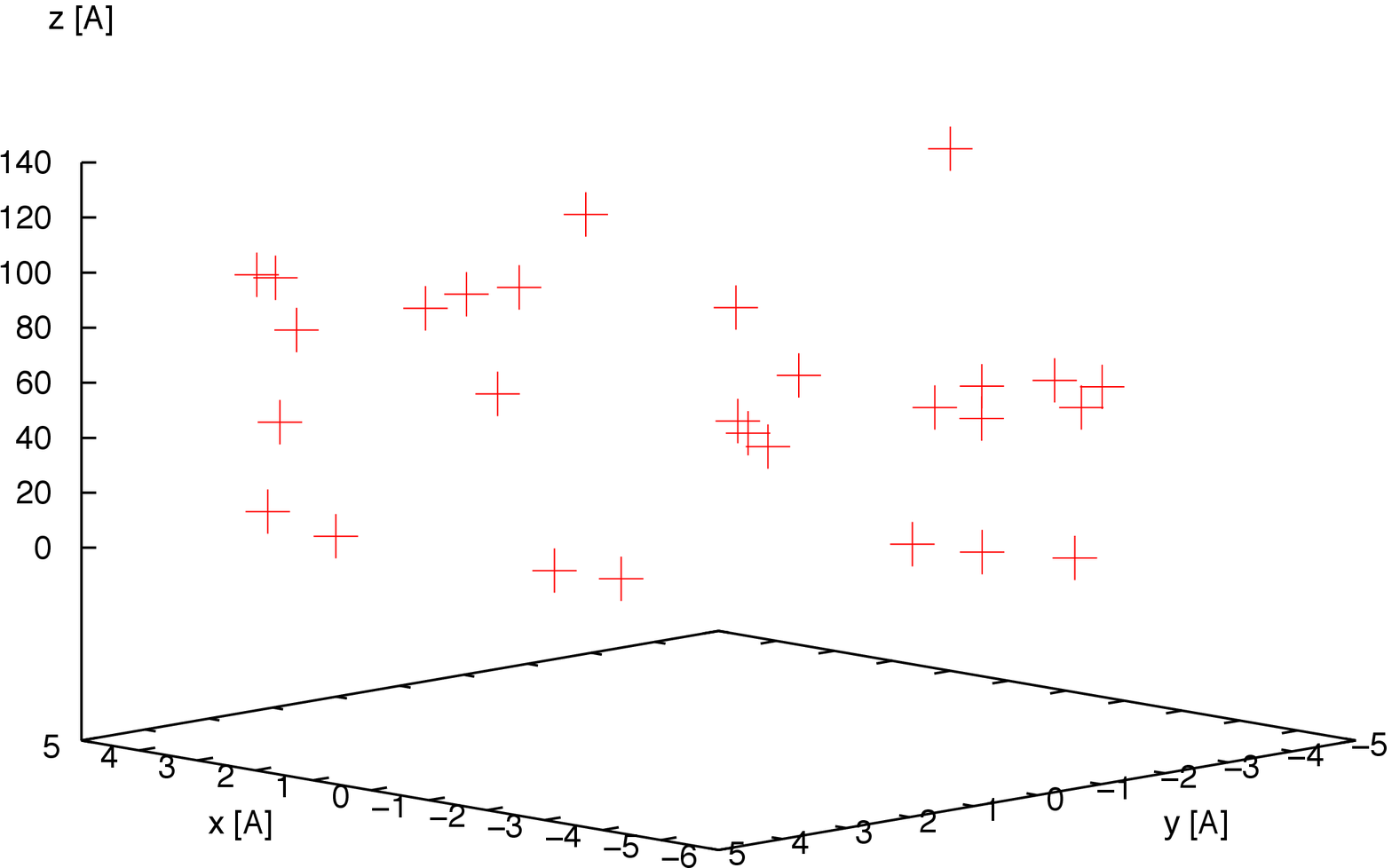}
\caption{\label{fig:dimer-samps}
H$_2$ spatial density distribution (at $T=1$~K, average density
$\bar\rho=1.46\times 10^{-3}$/\AA$^3$), for $N=28$ particles in a
pore of radius 7~\AA. Depicted are two individual Monte Carlo samples
of particle positions. (The scale of the vertical, longitudinal
axis is compressed relative to that of the transverse axes.) Note
the presence of dimers (H$_2$-H$_2$ pairs) in both plots, as well
as possible trimers.}
\end{figure}

\begin{figure}
\includegraphics[width=10cm,angle=-90]{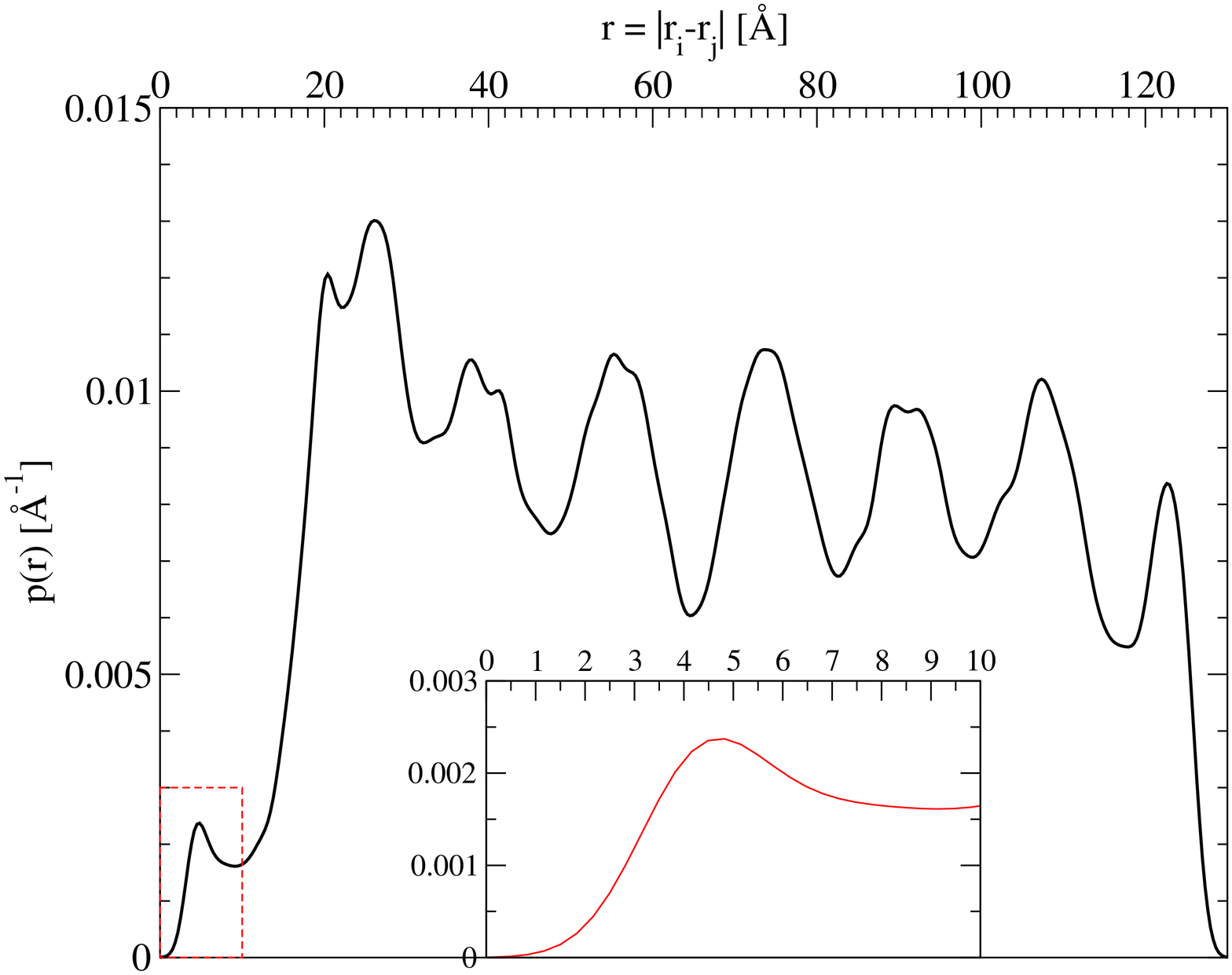}
\caption{\label{fig:pair-corr}
H$_2$ pair correlation function (at $T=1$~K, $\bar\rho=1.46\times
10^{-3}$/\AA$^3$), for $N=28$ particles in a pore of radius 7~\AA.
\textit{Inset:} An expanded view of the pair correlation function,
near the $r=4.8$~{\AA} peak, that indicates the presence of
dimerization.}
\end{figure}

A closer examination of samples drawn from Monte Carlo simulation,
Fig.~\ref{fig:dimer-samps}, reveals the presence of bound H$_2$-H$_2$
pairs, or ``dimers'', as well as possible bound triplets, or
``trimers''.  Such clustering is expected below a temperature set
by the H$_2$-H$_2$ binding energy. This visual evidence is corroborated
quantitatively by the pair correlation function derived from PIMC
simulation, Fig.~\ref{fig:pair-corr}, which shows an enhanced
probability for H$_2$ molecules to be separated by $r=4.8$~\AA. By
comparison, the classical equilibrium separation of a pair of H$_2$
molecules is $r=3.4$~\AA; the separation of quantum H$_2$ is greater
due to zero-point fluctuations, which reduce the binding energy to
about 3~K, roughly one-tenth of the well-depth of the pair
potential.\cite{ref31,ref37}

In a crude analysis of Fig.~\ref{fig:dimer-samps}, we estimate that
at $T=1$~K approximately equal numbers of H$_2$ molecules participate
in monomers (unpaired H$_2$) as dimers, i.e. a dimer/monomer ratio
of 0.5. This quantity can be roughly estimated by assuming that the
monomer$\Leftrightarrow$dimer process is in detailed balance,
implying that their respective chemical potentials, $\mu_1$ and
$\mu_2$, satisfy $\mu_2 = 2\mu_1$.  Assuming the H$_2$ to be an
ideal gas of $N$ molecules, the chemical potentials are related to
the individual populations ($N_i$) by
\begin{equation}
\begin{split} \mu_1 & = k_BT \ln\Bigl(\frac{N_1\lambda_1^D}{eV}\Bigr)\,,\\
\mu_2 & = k_BT \ln\Bigl(\frac{N_2\lambda_2^D}{eV}\Bigr) - \epsilon\,,
\end{split}
\end{equation}
where $\lambda \equiv \sqrt{2\pi\hbar^2/(mk_BT)}$ is the thermal
wavelength and $\epsilon$ is the binding energy per dimer.  These
relations, under the constraint $N=N_1+2N_2$, give a dimer/monomer
ratio
\begin{equation}
\frac{N_2}{N_1} = \frac{1+4x-\sqrt{1+8x}}{2(\sqrt{1+8x}-1)}\,,\qquad x \equiv 2^{D/2}\Bigl(\frac{\bar n}{n_Q}\Bigr) e^{\epsilon/(k_BT)}\,.\label{eq:mono-di}
\end{equation}
Here $\bar n=N/V$ is the mean 3D number density and $n_Q=1/\lambda^D$
is the quantum concentration.  For $N=28$ H$_2$ molecules at $T=1$~K
in a 3D volume $V=\pi R^2 L$, where $R=7$~{\AA} and $L=125$~\AA,
assuming their binding energy is $\epsilon=3$~K, Eq.~\ref{eq:mono-di}
predicts a dimer-monomer ratio of 8.5. This estimate exceeds the
ratio observed in PIMC simulation by an order of magnitude.  However,
the preceding back-of-the-envelope calculation is expected to have
limited accuracy since it ignores the quasi-2D geometry of adsorption,
the presence of internal rotational-vibrational excitational modes
of the dimer and the possibility of trimer, or larger, clusters of
bound molecules.

\subsection{Structure: semiclassical approximation}

As mentioned above, the axial and shell phases can be distinguished
in the classical, noninteracting case by the location of the potential
minimum ($r=0$ or $r>0$) for a given pore radius $R$.  Quantum
mechanically, one must resort to either solving the Schr\"odinger
equation and finding the peak probability from the radial wavefunction
(for the noninteracting case) or by carrying out a full path integral
Monte Carlo simulation (in the general interacting case).

\begin{figure}
\includegraphics[width=10cm,angle=-90]{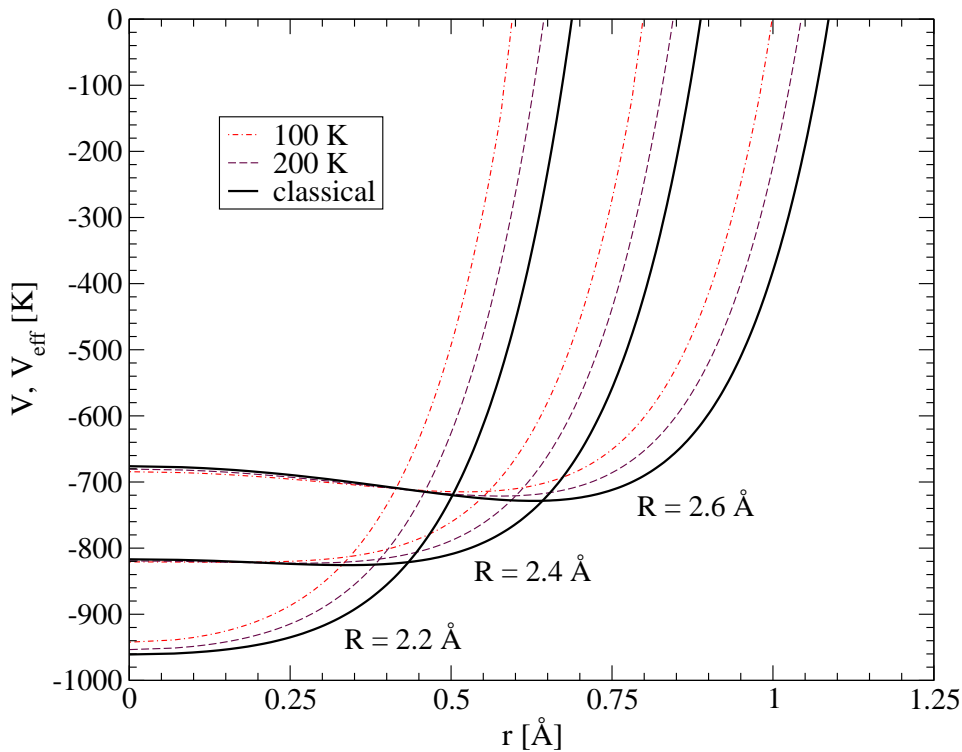}
\caption{\label{fig:Veff}
H$_2$-MgO pore radial effective potentials at $T=100$ and 200~K,
compared to the corresponding classical potentials, at pore radii
$R=2.2$, 2.4, and 2.6~\AA. Note that the quantum behavior described
by the effective potential approaches classical behavior in the
high temperature limit.}
\end{figure}

Besides their use in Monte Carlo simulations, the path integral
method also yields a semiclassical approximation to the density
that is almost as simple as inspection of the classical potential.
A variational approximation, originally due to Feynman, reduces the
quantum statistical path integral to the ordinary Boltzmann integral
of classical statistical mechanics, with the classical potential V
replaced by a semiclassical effective potential $V_\mathrm{eff}$
containing first-order quantum corrections to V.\cite{ref38}  This
potential is expressed as a weighted average of the real potential
over a distance comparable to the de Broglie thermal wavelength.
Thus, in this approximation, the axial phase of quantum H$_2$ occurs
when the minimum of the effective potential lies at $r=0$.  Feynman's
variational approximation yields a formula for $V_\mathrm{eff}$ as
a functional of the classical potential in the case of a noninteracting
quantum gas,
\begin{equation}
V_\mathrm{eff}(\mathbf{r}) = (2\pi\Lambda^2)^{-3/2} \int d^3\mathbf{r}^\prime\,V(\mathbf{r}^\prime) \exp{\Bigl[-\frac{1}{2}(\lvert\mathbf{r}-\mathbf{r}^\prime\rvert/\Lambda)^2\Bigr]},\qquad \Lambda \equiv \hbar/\sqrt{12mk_BT}\,.\label{eq:eff-pot}
\end{equation}
Here $m$ is the mass of a particle, in this case an H$_2$ molecule.
This has a high-temperature expansion
\begin{equation}
V_\mathrm{eff} = V + \frac{1}{2}\Lambda^2\nabla^2 V + O(\Lambda^4)\,. 
\end{equation}
The effective potential in Eq.~\ref{eq:eff-pot} is the convolution
of the classical potential with a Gaussian of width $\Lambda$, and
may be interpreted as a ``quantum smeared'' version of the classical
potential. In the high temperature (small $\Lambda$) limit, the
smearing vanishes, and the effective potential reproduces the
classical result. In the case of cylindrical symmetry, these equations
become
\begin{equation}
V_\mathrm{eff}(r) = \Lambda^{-2} \int_0^\infty dr^\prime\,V(r^\prime)\,r^\prime\,I_0\Bigl(\frac{rr^\prime}{\Lambda^2}\Bigr) \exp{\Bigl[-\frac{1}{2}(r^2+{r^\prime}^2)/\Lambda^2\Bigr]}\,,
\end{equation}
and
\begin{equation}
V_\mathrm{eff} = V + \frac{1}{2}\Lambda^2 (V^{\prime\prime}+V^\prime/r) + O(\Lambda^4)\,.
\end{equation}
Effective potentials for various $R$ and $T$ values are displayed
in Figure~\ref{fig:Veff}. As expected, $V_\mathrm{eff}$ approaches
$V$ as $T$ increases.  The main qualitative results concern the
location of the effective potential minimum. This energy is greater
than that of the classical potential minimum, due to the zero point
energy. It is also located closer to the axis, because the repulsive
hard wall of the potential at $r=R$ damps the wavefunctions of
quantum particles in the repulsive region.

\begin{figure}
\includegraphics[width=10cm,angle=-90]{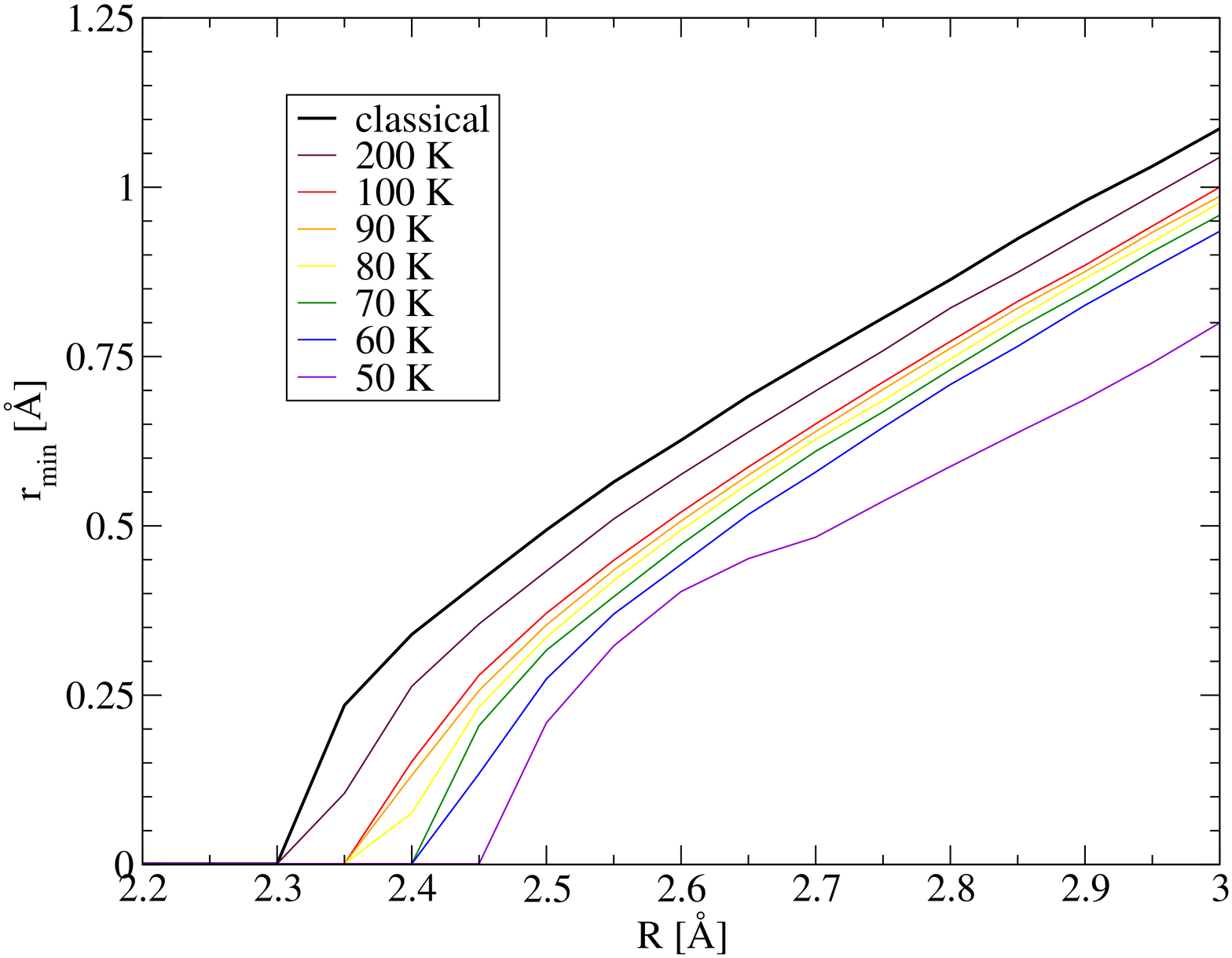}
\caption{\label{fig:rmin-effpot}
Minimum of the H$_2$-MgO pore radial effective potential,
$r_\mathrm{min}$, as a function of pore radius $R$, for various
temperatures (increasing from lower right to upper left), compared
to the minimum of the classical potential (upper leftmost curve).
The axial/shell transition occurs at the radius $R$ for which
$r_\mathrm{min}$ becomes nonzero.  This transition point for the
quantum gas is the same as the classical transition of $R=2.3$~{\AA}
when $T=200$~K, and increases to 2.45~{\AA} as the temperature of
the quantum gas decreases to $T=50$~K.  (The resolution of the plot
and therefore the precision of the estimated transition point is
calculated to within 0.05~\AA.)}
\end{figure}

Qualitatively, then, these semiclassical considerations imply that
a quantum gas has a greater tendency to be confined near the axis
than does a classical gas, agreeing with the preceding results
obtained by PIMC simulation.  Quantitatively, the semiclassical
approximation breaks down at temperatures above the $T=0.5$~K studied
by PIMC, but one may use it for qualitative purposes.  For example,
the location of the effective potential minimum at higher temperatures
($T>50$~K) is given in Figure~\ref{fig:rmin-effpot}. Note that the
classical potential minimum moves off axis near $R=2.30$--2.35~\AA,
while the effective potential minimum moves off axis near 2.35~{\AA}
at 100~K and near 2.45~{\AA} at 50~K.  By extrapolation of these
semiclassical predictions, it is plausible that the low $T$ axial-shell
transition occurs between 2.50 and 2.75~\AA, as found in the preceding
section by PIMC simulation at $T=0.5$~K.

\subsection{Heat capacity}

In accordance with the structural results of the preceding sections,
the heat capacity $C\equiv (\partial E/\partial T)_{N,V}$ of the
H$_2$ is expected to exhibit 1D behavior for small $R$, and 2D
behavior for larger $R$ pores. In the ideal gas approximation, for
small $R$, the single longitudinal degree of freedom should give a
dimensionless specific heat, or heat capacity per particle (in
$k_B=1$ units) of 1/2, in accordance with the equipartition
theorem.  An additional azimuthal degree of freedom is available
in the cylindrical shell phase of larger $R$, increasing $C/(Nk_B)$
to 1, given enough thermal energy to overcome the azimuthal excitation
energy gap. At very high $T$, the large radial excitation energy
gap will be exceeded, allowing an additional degree of freedom,
raising the specific heat to 3/2.

In the noninteracting case, the heat capacity can be calculated by
direct differentiation of the average energy, upon obtaining the
energy spectrum from numerical solution of the Schr\"odinger equation.
In the grand canonical ensemble, the average particle number and
energy of a Bose gas are given by summing the Bose-Einstein
distribution of states over all possible energies (weighted by the
density of states, or relative prevalence of each energy); this
includes an integration over the continuum of longitudinal energies
as well as a summation over the discrete spectrum of radial/azimuthal
transverse energies $\varepsilon_{n,\nu}$. These quantities are then
given by
\begin{align}
N(\beta,\mu) & = \sum N_\varepsilon = \sum_{n,\nu=0} g_\nu \int dE\,\frac{g(E)}{z^{-1}\exp[\beta(E+\varepsilon_{n,\nu})]-1}\,,\label{eq:N-nonint} \\
E(\beta,\mu) & = \sum E_\varepsilon = \sum_{n,\nu=0} g_\nu \int dE\,\frac{g(E)}{z^{-1}\exp[\beta(E+\varepsilon_{n,\nu})]-1}\,,\label{eq:E-nonint}
\end{align}
where $z$ is the fugacity [$\exp(\beta E)$], $\mu$ is the chemical
potential, $\beta=1/(k_B T)$ is the inverse temperature,
$g(E)=(L/\lambda)[\beta/(\pi E)]^{1/2}$ is the density of states
for a 1D (longitudinal) gas, and the degeneracy $g_\nu$ is 1 for
$\nu=0$ (no angular momentum) and 2 for $\nu>0$ [corresponding to
the $\pm$(counter-)clockwise azimuthal states].

Fixing the particle number $N$, the first equation is used to solve
$\mu$ in terms of particle number, $\mu(N)$, which is substituted
into the second equation to determine the average energy $E(T,N)$.
The specific heat $c=C/N$, or heat capacity per particle, is obtained
by a finite difference approximation to the derivative
$[\partial(E/N)/\partial_T]_{N,V}$.

\begin{figure}
\includegraphics[width=10cm,angle=-90]{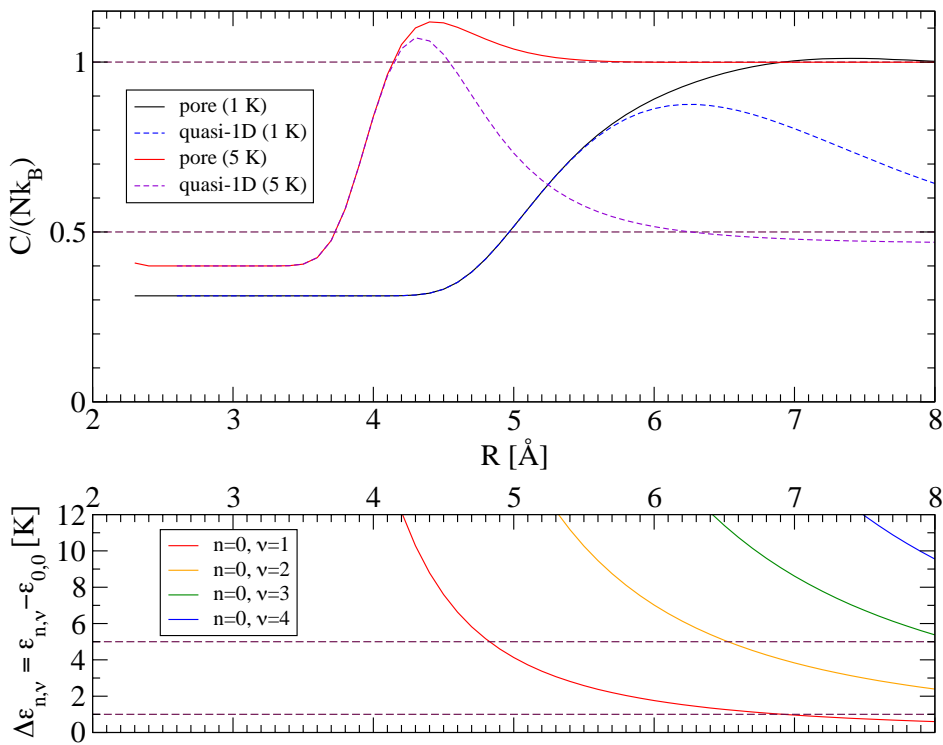}
\caption{\label{fig:cv-r}
\textit{Above:} Specific heat of noninteracting H$_2$ gas in a MgO
pore as a function of pore radius $R$, at fixed mean linear density
$N/L = 0.224$~\AA$^{-1}$ [mean 3D density $\bar\rho = N/L/(\pi
R^2)$], for $T=1$ and 5~K (right and left solid curves).  Contrasted
are the predictions of the quasi-1D model with an energy gap
$\varepsilon\equiv\varepsilon_{0,1}-\varepsilon_{0,0}$ determined
by the ground state and first azimuthal state (dashed curves).  (The
quantum numbers $n,\nu$ of $\varepsilon_{n,\nu}$ denote radial and
azimuthal excitations, respectively.)  The horizontal dashed lines
are the classical 1D and 2D limits of $\frac{1}{2}$ and 1, respectively.
\textit{Below:} The energy gap between the ground state
($\varepsilon_{0,0}$) and the first four azimuthal excitations of
the ground radial state ($\varepsilon_{0,\{1\ldots4\}}$, increasing
from left to right).  The horizontal dashed lines correspond to
energy gaps of 1 and 5~K.}
\end{figure}

The resulting dimensionless specific heat as a function of $R$, at
$T=1$ and 5~K, is depicted in Figure~\ref{fig:cv-r} (Also shown are
predictions for a quasi-1D approximation, in which only the azimuthal
ground and first excited states are considered, not discussed further
here.) The evolution of the specific heat from 1/2 for small $R$
to 1 for large $R$ is readily apparent. The small $R$ specific heat
is less than the classical prediction of $1/2$ because the temperatures
are low enough that the quantum system has not reached the classical
limit; it is closer to 1/2 for the more classical $T=5$~K case than
for the lower temperature, $T=1$~K, case.

The higher $T$ case approaches the 2D limit more quickly, since
more thermal energy is available to excite the higher azimuthal
modes that contribute the additional degree of freedom necessary
for 2D motion. Structurally, in the PIMC simulation at $T=0.5$~K,
the H$_2$ changed from an axial to a shell-like configuration for
$R<3$~\AA. The heat capacity, on the other hand, indicates a
transition from 1D to 2D behavior at larger $R\sim 6$~{\AA} at
$T=1$~K. Thus, the H$_2$ may behave thermodynamically as a 1D gas
even when it is structurally arranged as a 2D film. At pore sizes
of $R=4$~\AA, the pore is large enough for a shell configuration
to be energetically favorable, yet the azimuthal energy gap, which
shrinks with increasing $R$, is not yet large enough to permit
appreciable rotational motion leading to an enhanced heat capacity.
(The energy gaps of the first few azimuthal excited states are also
depicted in Fig.~\ref{fig:cv-r}.)

\section{Summary}
\label{sec:summary}

Results have been obtained for the $T=0$ properties of liquid $^4$He
and finite $T$ properties of liquid H$_2$. In both cases, the
behavior is seen to exhibit dimensional crossover as a function of
$R$, as shown in the density profiles and thermodynamic properties
(energies, chemical potentials and specific heat). A wide variety
of other properties, including detailed dependences on $N$ and $T$,
are amenable to further theoretical investigation. While we hope
to undertake some of these investigations in the future, they require
a substantial increase in computational resources and time.
Nevertheless, these are justified by recent experimental studies
of quantum fluids in and near carbon nanotubes and other porous
media.\cite{ref39,ref40,ref41,ref42}

\begin{acknowledgments}
We are grateful to Massimo Boninsegni for extensive discussions of
this problem and for providing the PIMC program. This research was
supported by grants from NSF DMR-0505160, from the Forum on
International Physics of the American Physical Society from Consejo
Nacional de Investigaciones Cientificas y T\'ecnicas of Argentina
(PID 5138/05) and from the University of Buenos Aires (X298). E.
S. H. wishes to thank the Department of Physics at Penn State for
its hospitality and generous support.
\end{acknowledgments}

\end{document}